\documentclass[letter]{aa} 
\usepackage{graphicx}
\usepackage{txfonts}
\usepackage{natbib}
\bibpunct{(}{)}{;}{a}{}{,} 
\usepackage{color}
\newcommand{\cmthree}{cm$^{-3}$}

\newcommand{\um}{$\mu$m}                                 
\newcommand{\lsun}{$L_{\odot}$}                          
\newcommand{\msun}{$M_{\odot}$}
\newcommand{\rsun}{$R_{\odot}$}

\newcommand{\mearth}{$M_{\oplus}$}

\newcommand{\mmoon}{$M_{\rm Moon}$}
\newcommand{\mjup}{$M_{\rm Jupiter}$}
\newcommand{\gapprox}{$\stackrel {>}{_{\sim}}$}   

\newcommand{\about}{$\sim$}                       
\newcommand{\powten}[1]{10$^{#1}$}
\newcommand{\qeri}{q$^{1}$\,Eri}
\newcommand{\epseri}{$\epsilon \, {\rm Eri}$}

\newcommand{\amin}{$^{\prime}$}                   
\newcommand{\asec}{$^{\prime \prime}$}
\newcommand{\adeg}{$^{\circ}$}

\newcommand{\radot}[4]{\mbox{#1$^{\rm h}$#2$^{\rm m}$#3$\stackrel{\rm s}
{_{\bf\cdot}}$#4}}  

\newcommand{\decdot}[4]{\mbox{#1$^{\circ}$ #2$^{\prime}$ #3$\stackrel {\prime 
\prime}{_{\bf \cdot}}$#4}}

\newcommand{\asecdot}[2]{\mbox{#1$\stackrel {\prime \prime}{_{\bf \cdot}}$#2}}

\begin{document}
   \title{q$^{1}$\,Eri: a solar-type star with a planet and a dust belt\thanks{Based on observations with APEX, Llano Chajnantor, Chile.}}


   \author{R. Liseau\inst{1}
          \and
          C. Risacher\inst{2}
          \and
           A. Brandeker\inst{3}
	  \and
          C. Eiroa\inst{4}
	  \and
          M. Fridlund\inst{5}
	  \and
          R. Nilsson\inst{3}
	  \and
          G. Olofsson\inst{3}
	  \and
          G.L. Pilbratt\inst{5} 
	  \and	\\
	   P. Th\'ebault\inst{3,\,6}
          }

   \offprints{R. Liseau}

   \institute{Onsala Space Observatory, 
	      Chalmers University of Technology, S-439 92 Onsala, Sweden\\
              \email{rene.liseau@chalmers.se}
         \and
             European Southern Observatory, Casilla 19001, Santiago 19, Chile\\
             \email{crisache@eso.org}
        \and
             Stockholm Observatory, AlbaNova University Center,
             Roslagstullsbacken 21, SE-106 91 Stockholm, Sweden\\
	     \email{alexis@astro.su.se,ricky@astro.su.se,olofsson@astro.su.se,philippe.thebault@obspm.fr}
         \and
             Departamento de F\'{i}sica Te\'{o}rica, C-XI, Facultad de Ciencias,
             Universidad Aut\'{o}noma de Madrid, Cantoblanco, 28049 Madrid, Spain\\
             \email{carlos.eiroa@uam.es}
         \and
	     ESA Astrophysics Missions Division, 
             ESTEC, PO Box 299, NL-2200 AG Noordwijk, The Netherlands\\
             \email{malcolm.fridlund@esa.int,gpilbratt@rssd.esa.int}
         \and	 
	     LESIA, Observatoire de Paris, F-92195 Meudon Principal Cedex, France 
	}

   \date{Received ; accepted }


  \abstract
   {Far-infrared excess emission from main-sequence stars is due to dust produced by orbiting minor bodies. In these disks, larger bodies, such as planets, may also be present and the
understanding of their incidence and influence currently presents a challenge.}
   {Only very few solar-type stars exhibiting an infrared excess and harbouring planets are known to date. Indeed, merely a single case of a star-planet-disk system has previously been detected at submillimeter (submm) wavelengths. Consequently, one of our aims is to understand the reasons for these poor statistics, i.e., whether these results reflected the composition and/or the physics of the planetary disks or were simply due to observational bias and selection effects. Finding more examples would be very significant.}
   {The selected target, \qeri, is a solar-type star, which was known to possess a planet, \qeri\,b, and to exhibit excess emission at IRAS wavelengths, but had remained undetected in the millimeter regime. Therefore, submm flux densities would be needed to better constrain the physical characteristics of the planetary disk. Consequently, we performed submm imaging observations of \qeri.}
   {The detected dust toward \qeri\ at 870\,\um\ exhibits the remarkable fact that the entire SED, from the IR to mm-wavelengths, is fit by a single-temperature blackbody function (60\,K). This would imply that the emitting regions are confined to a narrow region (ring) at radial distances much larger than the orbital distance of \qeri\,b, and that the emitting particles are considerably larger than some hundred micron. However, the 870\,\um\ source is extended, with a full-width-half-maximum of roughly 600\,AU. Therefore, a physically more compelling model also invokes a belt of cold dust (17\,K), located at 300\,AU from the star and about 60\,AU wide.}
   {The minimum mass of 0.04\,\mearth\ (3\,\mmoon) of 1\,mm-size icy ring-particles is considerable, given the stellar age of \gapprox\,1\,Gyr. These big grains form an inner edge at about 25\,AU, which may suggest the presence of an unseen outer planet (\qeri\,c). }

   \keywords{Stars: individual: q$^{1}$\,Eri (HD\,10647) -- 
             Stars: planetary  systems: planetary disks -- 
             Stars: planetary systems: formation
               }

   \maketitle
%

\section{Introduction}

During the end stages of early stellar evolution, dusty debris disks are believed to be descendents of gas-rich protoplanetary disks. These had been successful to varying degrees in building a planetary system. What exactly determines the upper cut-off mass of the bodies in individual systems, and on what time scales, is not precisely known. However, the presence of debris around matured stars is testimony to the action of orbiting bodies, where a large number of smaller ones are producing the dust through collisional processes and where a small number of bigger bodies, if any, are determining the topology (disks, rings and belts, clumps) through gravitational interaction. The time evolution of the finer debris is believed to be largely controlled by non-gravitational forces, though. By analogy, many debris disks are qualitatively not very different from the asteroid and Kuiper belts and the zodiacal dust cloud in the solar system \citep{mann2006}. 

For solar-type stars on the main-sequence, which are known to exhibit infrared excess due to dust disks, one might expect, therefore, a relatively high incidence of planetary systems around them. Surveying nearly 50 FGK stars with known planets for excess emission at 24\,\um\ and 70\,\um, \citet{trilling2008} detected about 10-20\% at 70\,\um, but essentially none at 24\,\um, implying that these planetary disks are cool ($<100$\,K) and large ($> 10$\,AU). However, in general, the conjecture that the infrared excess arises from disks lacks as yet observational confirmation due to insufficient spatial resolution. In fact, until very recently, there was only one main-sequence system known that has an extended, resolved disk/belt structure and (at least) one giant planet, viz. \epseri, a solar-type star at the distance of only three parsec \citep{greaves1998,greaves2005}. Its planetary companion, \epseri\,b, has been detected indirectly by astrometric and radial velocity (RV) methods applied to the star \citep{hatzes2000,benedict2006}, whereas attempts to directly detect the planet have so far been unsuccessful \citep{itoh2006,janson2007}.

As its name indicates, the object of the present study, \qeri, happens to belong to the same celestial constellation of Eridanus, albeit at a larger distance ($D=17.35 \pm 0.2$\,pc) and is, as such, unrelated to \epseri. The planet was discovered with the RV technique \citep[for a recent overview, see][]{butler2006}. These RV data suggest that the semimajor axis of the Jupiter-mass planet \qeri\,b is about 2\,AU (Table\,\ref{star}). It seems likely that regions inside this orbital distance have been largely cleared by the planet, whereas outside the planetary orbit, substantial amounts of material might still be present. 

In fact, IRAS and ISO data were suggestive of significant excess radiation above the photospheric emission at wavelengths longward of about 20\,\um. \citet{zuckerman2004} interpreted these data in terms of dust in a disk at the orbital distance of 30\,AU and at a temperature of about 55\,K. \citet{chen2006} fitted the far-infrared emission with the corresponding values of 20\,AU and 70\,K, respectively. \citet{trilling2008} derived 20\,AU and 60\,K. In their entire sample of more than 200 stars, \qeri\ (=HD\,10647) has by far the highest 70\,\um\ excess.

At mm-wavelengths, \citet{schutz2005} failed to detect the disk and assigned an upper limit to the dust mass of 6\,\mmoon. This is unsatisfactory, as the proper characterization of the dust around \qeri\ would require valid long wavelength data. In the following, observations of \qeri\ at 870\,\um\ are described and their implications discussed.

\section{Observations and Data Reductions}

APEX, the Atacama Pathfinder EXperiment, is a 12\,m diameter submillimeter telescope situated at an altitude of 5100\,m on the Llano Chajnantor in northern Chile. The telescope is operated by the Onsala Space Observatory, the Max-Planck-Institut f\"ur Radioastronomie, and the European Southern Observatory.

The Large Apex BOlometer CAmera \citep[LABOCA,][]{siringo2007} is a multi-channel bolometer array for continuum observations with 60\,GHz band width and centered on the wavelength of 870\,\um. The array, having a total field of view of 11\amin, is spatially undersampled and we therefore adopted spiral pattern observing as the appropriate technique \citep{siringo2007}. This procedure results in fully-sampled maps with a uniform noise distribution over an area of about 8\amin. During the nights of August\,1-4, 2007, we obtained 32 such individual maps, for about 7.5\,min each with central coordinates RA=\radot{01}{42}{29}{32} and Dec=\decdot{$-53$}{44}{27}{0} (J2000).  The LABOCA beam width at half power (HPBW) is \asecdot{18}{6} $\pm$ \asecdot{1}{0}. We focussed LABOCA on the planet Jupiter and the rms-pointing accuracy of the telescope was 3\asec\ to 4\asec.

\begin{figure}[t]
  \resizebox{\hsize}{!}{
  \rotatebox{270}{\includegraphics{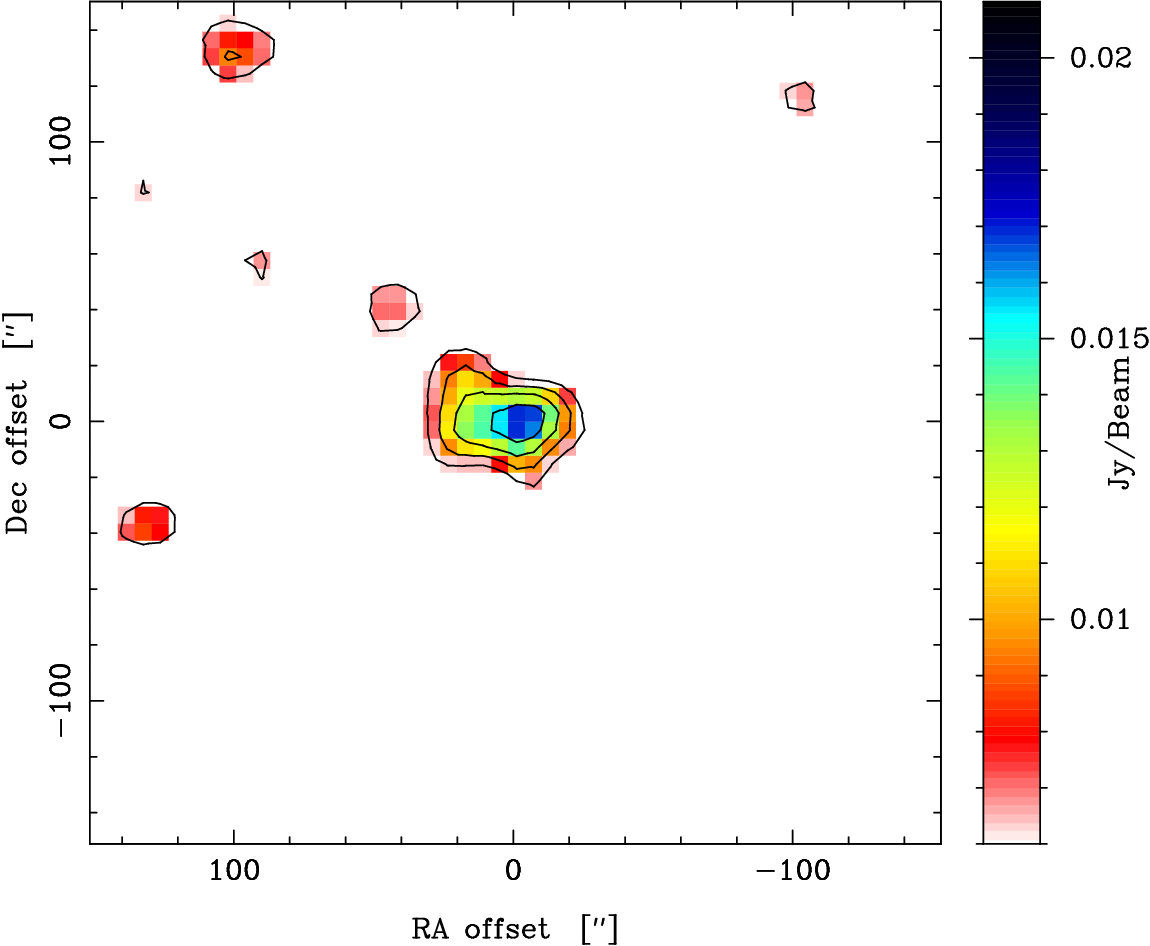}}
  }
  \caption{\qeri\ observed at 870\,\um\ with the submm camera LABOCA at the APEX telescope (HPBW\about18\asec). Within the positional accuracy, the star is at the origin of the image (see Table\,\ref{parameters}, referring to \radot{01}{42}{29}{32}, \decdot{$-53$}{44}{27}{0}, J2000.0) and offsets are in arcsec. The lowest contour corresponds to $2 \sigma$ of the rms noise of the flux density, $F_{\nu}$, and increments are in steps of $1 \sigma$. The colour coding, in units of Jy/beam, is shown by the scale bar to the right of the image, which has been smoothed with a circular 18\asec\ Gaussian. At the distance of the star, the space between two tick marks (=20\asec) corresponds to 350\,AU.
  }
  \label{obs}
\end{figure}

We reduced the data with the BoA software \citep{siringo2007}, which included flat fielding, baseline removal, despiking and iteratively removing the sky noise, and filtering out the low frequencies of the $1/f$-noise, with the cut-off frequency corresponding to several arcminutes. The software also accounts for the map reconstruction and the absolute calibration, using the opacities determined from numerous skydips (zenith opacities were in the range 0.1 to 0.3) and observations of the planets Uranus and Mars. The final result is an rms-noise-weighted average map (Fig.\,\ref{obs}).

\begin{table}
\begin{flushleft}
 \caption{\label{star} Physical properties of the star and its planet
$^{\star}$}
\begin{tabular}{ll}
  \hline
  \noalign{\smallskip}
Parameter                                               & Value   \\
  \noalign{\smallskip}
\hline \\                                 
{\bf The star \qeri}                                    &         \\
  \noalign{\smallskip}   
Distance, $D$                                           & 17.35\,pc \\
Spectral type and luminosity class                      & F8-9\,V  \\    
Effective temperature, $T_{\rm eff}$                    & 6100\,K  \\   
Luminosity, $L_{\rm star}$                              & 1.2\,\lsun  \\   
Surface gravity, $\log {g}$                             & 4.4 (in cm s$^{-2}$)\\   
Radius, $R_{\rm star}$                                  & 1.1\,\rsun  \\  
Mass $M_{\rm star}$                                     & 1.1\,\msun  \\  
Metallicity, [Fe/H]                                     & $-0.08$  \\ 
Age                                                     & $(>1-2)$\,Gyr \\ 
  \noalign{\smallskip}
{\bf The planet \qeri\,b}                               &         \\
  \noalign{\smallskip}    
Period, $P$                                             & $2.75 \pm 0.15$\,yr \\
Semimajor axis, $a_{\rm orbit}$                         & $2.0 \pm 0.2$\,AU \\
Eccentricity, $e$                                       & $0.2 \pm 0.2$     \\
Mass, $M \sin{i}$                                       & $0.9 \pm 0.2$\,\mjup\\
  \noalign{\smallskip}
  \hline
  \end{tabular}
\end{flushleft}
$^{\star}$ See \citet{butler2006} and references cited in the text. 
\end{table}

\section{Results}

The final product of the reduction process is the 870\,\um\ image presented in Fig.\,\ref{obs}, which shows the central $5^{\prime} \times 5^{\prime}$ of the LABOCA map. The peak flux in the map is found at the position of \qeri\ and a few other pointlike features of low intensity are also present, one of which is close to the star. If not merely noise, these low signals could be due to extragalactic background sources, as the displayed number density is consistent with that observed elsewhere \citep[e.g.,][]{lagache2005,bertoldi2007,ivison2007}. Other, complementary observations (e.g., optical, IR, X-rays, radio interferometry) would be required for their identification.

The derived flux densities of \qeri\ are provided in Table\,\ref{parameters}, which presents the results from fitting the data to a two-dimensional Gaussian function. The indicated errors are formal fit errors only, based on $1 \sigma$ rms values. In individual cases, e.g., {\it pa}, realistic errors could be twice as large. The error on the integrated flux density in Table\,\ref{parameters} also includes an uncertainty of 10\% in the absolute calibration.

The 870\,\um\ source is at best only marginally resolved in the North-South direction (formal fit result is 23\asec\ $\pm$ 1\asec), whereas it is clearly elongated in approximately the East-West direction (37\asec\ $\pm$ 2\asec). At the distance of \qeri, this corresponds to a disk diameter of 640\,AU and assuming a circular shape, these disk dimensions yield an inclination with respect to the line of sight of $i \ge 52$\adeg, not excluding the possibility that the disk is seen essentially edge-on ($i$ close to 90\adeg). The vertical disk scale height is undetermined. 

\section{Discussion}

\subsection{Physical conditions and the age of the system}

The spectral type of \qeri\ is slightly earlier than that of the Sun \citep[F8-9~V,][]{nordstrom2004,decin2000,decin2003,zuckerman2004,chen2006}, with the effective temperature being bracketed by the extremes 6040\,K \citep{nordstrom2004} and 6260\,K \citep{chen2006}
, with the mean of 6150\,K, i.e., essententially the value given by \citet[][6105~K]{butler2006} (see Table\,\ref{star}).

Literature estimations of likely ages for \qeri\ span the range 0.3 to 4.8\,Gyr \citep[with an entire range of 0.0 to 7.0\,Gyr,][]{decin2000,zuckerman2004,decin2003,nordstrom2004,chen2006}. However, the level of chromospheric activity ($\log{R^{\prime}_{\rm HK}}=-4.7$) suggests an age of 1.9\,Gyr \citep[see Eq.\,15 of][]{wright2004}. The star has also been detected in X-rays with ROSAT ($\log{L_{\rm X}}=28.3$, J.\,Sanz, private communication), yielding 1.2\,Gyr \citep{ribas2005,guinan2007}. This value is also consistent with the stellar rotation period of about 10\,days \citep[uncorrected for $\sin i$;][]{ecuvillon2007}. It is clear that the star is definitely on the main-sequence and that the age of the system likely exceeds \powten{9}\,yr.

\begin{figure}[t]
  \resizebox{\hsize}{!}{
  \rotatebox{00}{\includegraphics{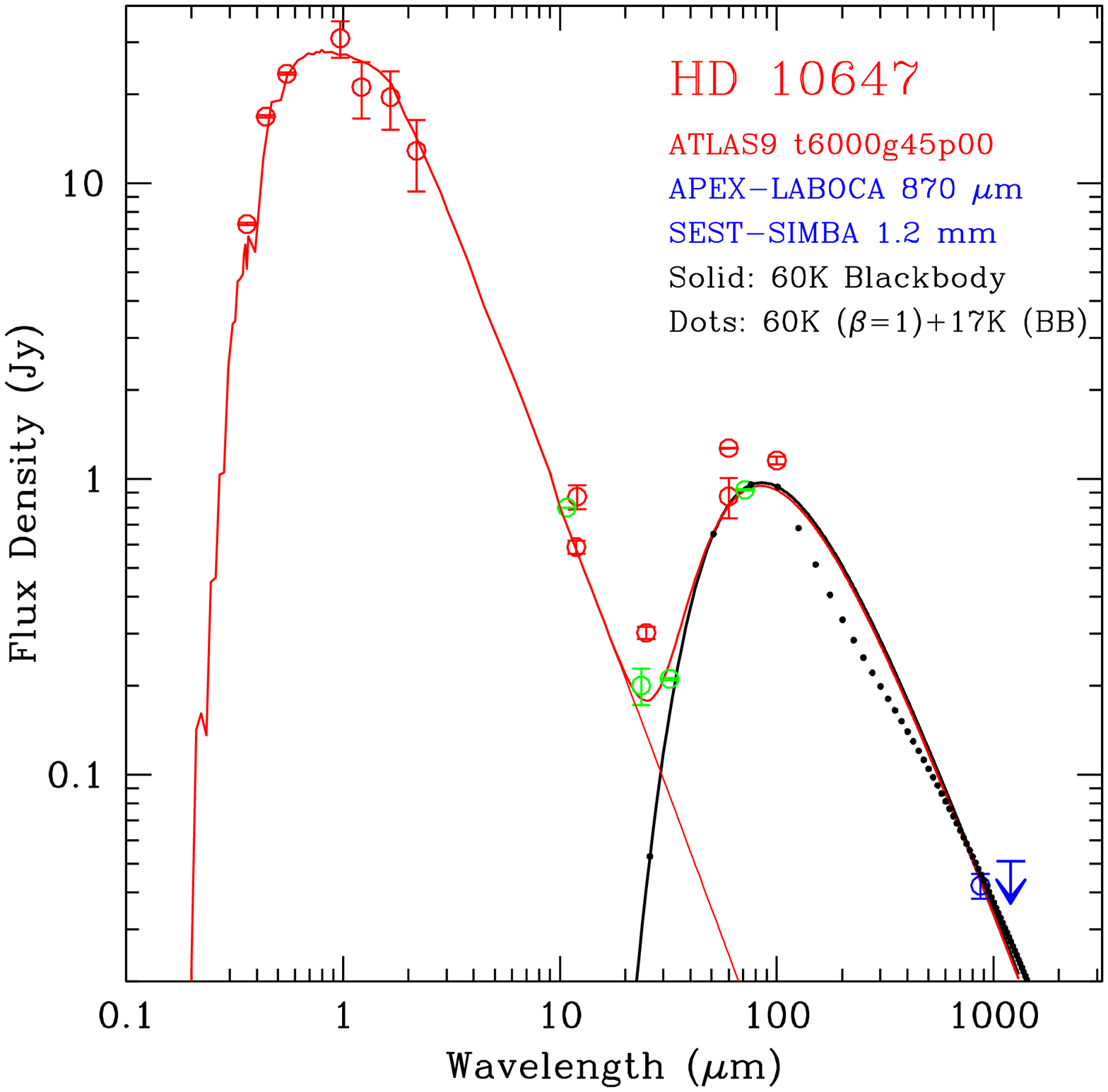}}
  }
  \caption{The fit to the SED of \qeri\ (HD\,10647) is shown from 1000\,\AA\ to beyond 1\,mm in red. The stellar photosphere is represented by a Kurucz ATLAS\,9 model atmosphere and the excess emission by a single temperature blackbody curve (in black). Also, shown in red, are Simbad, TIMMI\,2 11.9 \,\um, and IRAS FSC data, in addition to ISO 60\,\um\ \citep{decin2000}. Spitzer data have been given higher weight and are shown in green, viz. 8.5-13\,\um\ and 30-34\,\um\ \citep{chen2006}, and 24\,\um\ and 70\,\um\ \citep{trilling2008}, respectively. The SEST-SIMBA $3 \sigma$ upper limit at 1.2\,mm \citep{schutz2005} and our APEX-LABOCA point at 870\,\um\ are shown in blue. The black dots refer to the combined 60 K modified blackbody ($\beta =1$) and 17\,K blackbody (see the text).
  }
  \label{SED}
\end{figure}

\begin{table}
\begin{flushleft}
 \caption{\label{parameters} Physical properties of the \qeri\ dust system}
\begin{tabular}{ll}
  \hline
  \noalign{\smallskip}
Parameter                                               & Value   \\
  \noalign{\smallskip}
\hline \\                                 
Peak offset$^a$, ($\Delta \alpha,\,\Delta \delta$) 	& (+4\asec, +3\asec), (error: $\pm 4$\asec) \\
Peak flux density$^a$, $F_{\nu}$(0, 0), $\lambda=870$\,\um& $(16.2 \pm 0.8)$\,mJy/beam \\
Integrated flux density, $\int\!F_{\nu} d\alpha d\delta$& $(39.4 \pm 4.1)$\,mJy, $F_{\nu} \ge 2 \sigma$ \\
Major axis$^a$ (FWHM)					& 37\asec $\pm$ 2\asec\ (640 $\pm$ 35)\,AU  \\    
Position angle$^a$, $pa$				& 55\adeg\ $\pm$ 4\adeg\ (north over east) \\
Minor axis$^a$ (FWHM)					& 23\asec\ $\pm$ 1\asec \\  
Inclination angle, $i$					& $\ge 52$\adeg\ (90\adeg\,=\,edge-on) \\
Fractional luminosity, $L_{\rm bb}/L_{\rm star}$	& $1.1 \times 10^{-4}$ \\
Inner (outer)$^b$ Temperature, $T_{\rm bb}$		& 60\,K (17\,K) \\
Inner (outer)$^b$ Radius, $r_{\rm bb}$			& 25\,AU (300\,AU) \\
Inner (outer)$^b$ Width, $\Delta r_{\rm bb}$		& 0.02\,AU (60\,AU) \\
Inner (outer)$^b$ Minimum mass$^{c}$, $M_{\rm dust}$	& 0.04\,\mearth\ (0.15\,\mearth) \\
  \noalign{\smallskip}
  \hline
  \end{tabular}
\end{flushleft}
$^{a}$ Two-dimensional Gaussian fits with $1 \sigma$ {\it formal} fitting uncertainties. \\
$^{b}$ An outer dust belt is implied by the extent of \qeri\ at 870\,\um. \\
$^{c}$ $\kappa_{10^{11.5}\,{\rm Hz}}=2$\,cm$^2$\,g$^{-1}$ ($\rho=1.18$\,g\,\cmthree, $a_{\rm max}=1$\,mm, $n(a) \propto a^{-3.5}$). 
\end{table}

\subsection{The nature of the emitting particles}

The absence of spectral features in the 10 to 30\,\um\ region suggests that the dust grains are considerably larger than 10\,\um\ \citep{chen2006,schutz2005}. Remarkably, the spectral energy distribution (SED) of the excess emission can be fit by a single-temperature blackbody of 60\,K, from the infrared to the submm/mm regime (see Fig.\,\ref{SED}). The blackbody character is determined by the LABOCA flux and independent of the relative weights assigned to the mid- and far-infrared data. The radial distance from the central star, at which a grain has attained thermal equilibrium, is approximately given by $[(1-A)/(16 \pi\,\epsilon\,\sigma)\,(L_{\rm star}/T^4)]^{1/2}$, where $A$ and $\epsilon$ are the integrated reflectivity and emissivity, respectively. For a blackbody this reduces to $r_{\rm bb} = (R_{\rm star}/2)\,(T_{\rm eff}/T_{\rm bb})^2$, which for $T_{\rm bb}=60$\,K, yields a minimum distance of 25\,AU for the \qeri\ dust (Table\,\ref{parameters}). Taken at face value, this would mean that the range in dust temperatures is very limited: single values lower than 50\,K or as high as 100\,K can be excluded (without giving higher weight to the Spitzer data, $T_{\rm bb}$ becomes closer to 70\,K). Therefore, $r_{\rm bb}$, is determined to better than within a factor of two. For unit filling factor, the blackbody emitting regions would appear to be confined to a very narrow ring-like structure (see Table\,\ref{parameters}). 

The fractional luminosity is $1 \times 10^{-4}$ and  the emission is optically thin. The blackbody fit also implies that the emitting particles have sizes largely in excess of 100\,\um\ ($2 \pi a > \lambda$) and that these grains have grey opacities in the infrared to submm, i.e., $\kappa \neq \kappa(\lambda)$. Given the available evidence, it is not possible, however, to tell the actual sizes of the particles or their absolute opacities. 

Some insight may be gained from the work of \citet{miyake1993}, who explored the optical properties of dust that produces small values of the opacity index, and presented opacities over a broad range in frequency and particle size. Maximum opacity, max\,$\kappa \sim 2$\,cm$^2$\,g$^{-1}$, was found for the size $a_{\rm max}=1$\,mm at $\nu = 10^{11.5}$\,Hz ($\lambda \sim 1$\,mm) and for larger particles, $\kappa$ decreases rapidly (as $\sim 1/\sqrt{a_{\rm max}}$). This assumes compact spheres of density $\rho=1.18$\,g\,\cmthree\ (well-mixed silicates and water ice) and being distributed in size according to $n(a) \propto a^p$, with $p=-3.5$. The adopted density is consistent with values determined for Kuiper Belt objects \citep[][and references therein]{grundy2007}. The value of $\kappa_{1\,{\rm mm}}$ is not strongly dependent on $p$, as long as $-4 \le p \le -2$ \citep{miyake1993}. In general, these results are in agreement with other work \citep[e.g.,][]{krugel1994,stognienko1995}.

For this maximum value of $\kappa$, the 870\,\um\ flux density yields a minimum mass $M_{\rm dust} = F_{\nu}\,D^2/\kappa_{\nu}\,B_{\nu}(T_{\rm dust}) \ge 3$\,\mmoon\ (0.04\,\mearth, see Table\,\ref{parameters}). This minimum mass is larger than the 'blackbody mass' one would infer from the effective area of the blackbody ($6.85 \times 10^{26}$\,cm$^2$) and for the same $a$ and $\rho$, but consistent with the (re-scaled) result of \citet{schutz2005}. 

The fact that the 870\,\um\ source appears linearly resolved, speaks against the narrow ring scenario, and the existence of a more massive and colder belt ($T=17$\,K at $r=300$\,AU, say) can at present not be excluded (see Fig.\,\ref{SED}). The relative width of such a cold belt would seem less implausible, viz. $\Delta r/r \ge 0.2$ and, hence, its physical width would be at least 60\,AU. Also, the equilibrium temperature of the dust would be esssentially constant. With the same parameters as before, the mass would scale simply as the ratio of the  temperatures, yielding 13\,\mmoon. Of course, at shorter wavelengths, the spectrum would have to be steeper than the blackbody SED, implying values of $\beta > 0$, where $\beta$ parameterizes the frequency dependence of the opacity, i.e., $\kappa_{\nu} \propto \nu^{\,\beta}$. Together with the 17\,K blackbody, the SED can be fit with a modified 60\,K blackbody with $\beta=1.0$ longward of 100\,\um\ (Fig.\,\ref{SED}). This model would physically be more attractive, but by its ad hoc nature would be of course not unique, and better constraints on the physical parameters would require a better sampling of the SED. 

\subsection{Relation to the planet \qeri\,\,b}

An index of the order of $-3.5$, used by \citet{miyake1993}, could indicate that the size distribution resulted from a collisional cascade \citep[for a discussion, see][]{thebault2007}. The observed absence of small debris ($a$ of the order of 1\,\um\ or smaller) in the \qeri\ disk suggests that its production has ceased and that it had diminished on a short time scale compared to the age of the system. Remarkably, the density parameter of \citet{wyatt2005} has a value that, given the age of \qeri, is atypically large ($\eta_0 \sim 1000$ for the correct form of Eq.\,7), which may mean that the observed absence of warm material is caused by radiation pressure blow-out, rather than by the action of a planet, hindering the migration inward toward the star. Anyway, at 2\,AU distance, the Jupiter-size planet \qeri\,b could hardly have had any influence on particles out to the innermost edge at 25\,AU and on orbits far beyond that \citep{wyatt2005}. The existence of this belt of large grains may point to the presence of another major planet. It would therefore seem important to verify or to disprove the existence of \qeri\,c.

\section{Conclusions}

Below, our main conclusions from this work are summarized:

\begin{itemize}
\item[$\bullet$] Observations of the solar-type star \qeri\ and its planet \qeri\,b at 870\,\um\ revealed a source with peak emission at the position of the star. The source appears extended to the LABOCA beam, i.e., elongated in roughly the East-West direction (640\,AU) but essentially unresolved in the perpendicular direction. 
\item[$\bullet$] At an age exceeding 1\,Gyr, the fractional luminosity of the infrared excess is very high ($\ge 10^{-4}$). The entire SED of this excess emission, extending from \gapprox\,20\,\um\ to 1\,mm, is fit by a single-temperature blackbody ($T_{\rm bb}=60$\,K).
\item[$\bullet$] This would imply that the emitting regions are located at about 25\,AU from the star and in addition, very limited in spatial extent (ring-like). Exhibiting a grey opacity over the entire wavelength range, the emitting particles must be large ($>>100$\,\um). Using the theoretically derived maximum grain opacity for 1\,mm-size icy particles, we estimate a minimum mass of the dust belt of 0.04\,\mearth.
\item[$\bullet$] It seems highly unlikely that the planet \qeri\,b at 2\,AU would be responsible for the clearing of the region from small dust particles interior to 25\,AU, but may hint at the existence of another planet. 
\item[$\bullet$] Taking the observed extent of the 870\,\um\ source into account leads to an emission model in which an outer cold (17\,K) dust belt needs to be included. This belt would be centered on the radial distance of 300\,AU and have a width of 60\,AU. This belt is inclined at $i \ge 52$\adeg, possibly viewed at an angle close to edge-on.
\item[$\bullet$] The extreme character of the debris disk around the relatively old star-planet system \qeri\ provides yet another example of the large diversity of such disks.
\end{itemize}

\acknowledgement{ 
We are indebted to the staff at the APEX facility in Sequitor and at the telescope site for their enthusiastic and skillful support during our observing run. The thoughtful comments by the referee were much appreciated.
}

\end{document}